\documentstyle[12pt]{article}

\begin{document}
\begin{center}
{\LARGE{\bf On the Structure of Quantum Gauge\\
            Theories with External Fields}}\\
\vspace{0.7 cm}
{\large S.~Falkenberg and  B.~Geyer}\footnote{E-mail:
geyer@rz.uni-leipzig.de}\\
{\small{\it Center of Theoretical Sciences, and Institute
of Theoretical Physics, \\Leipzig University,
Augustusplatz 10--11, D-04109 Leipzig, Germany}}\\

{\large P.~Lavrov}\footnote{E-mail: lavrov@tspi.tomsk.su}\\
{\small{\it Tomsk State Pedagogical University, Tomsk 634041, Russia}}\\

and\\

{\large P.~Moshin}\\
{\small
{\it Tomsk State University, Tomsk 634050, Russia}}\\
\end{center}
\vspace{0.1 cm}
\begin{quotation}
 \small\noindent
 We consider generating functionals of Green's functions with external
 fields in the framework of BV and BLT quantization schemes for general
 gauge theories. The corresponding Ward identities are obtained, and the
 gauge dependence is studied.
\end{quotation}
\section{Introduction}
 The Ward identities (i.e. relations between Green's functions resulting
 from initial classical invariance) are
 the basic means providing insight into the quantum structure of gauge
 theories. Thus, they are indispensable for the proof of gauge-invariant
 renormalizability of general gauge theories in both the BV \cite{1} and
 BLT \cite{2} quantization approaches. Also based on the use of the Ward
 identities was the investigation of gauge dependence, originally, in
 Yang-Mills theory \cite{3} and, afterwards, in general gauge
 theories \cite{4,5} (this equally refers to theories with composite
 fields \cite{6,7}). At the same time, in quantum general gauge theories
 (both non-renormalized and renormalized ones), the Ward identities
 underlie the proof of the existence of Noether charge operators \cite{8}
 with the algebraic properties required for the analysis of unitarity
 conditions \cite{9}.

 Considerably less appreciated, however, is the effect which Ward identities
 may have on the study of quantum gauge theories with external fields
 (although examples of such theories, arousing interest in the
 corresponding Ward identities, recently have appeared: see, for instance,
 \cite{10}).

 The concept of external fields is basically utilized in the
 study of theories (notably those coupled to gravity) that are faced with
 difficulties in performing integration over a part of the variables
 in the path integral. The well-known attempts to handle such problems in
 quantum field theory consist in lifting functional integration from these
 variables (normally, a part of the initial classical fields), which are
 then treated as external. Note that so far this has been considered as
 the only reliable approach --- in the absence of a consistent theory of
 quantum gravity --- permitting to take into account such gravitational
 effects that pertain to the GUT epoch of the Universe, e.g. to
 incorporate the effects of gravity into the early Universe cosmology and
 to address the numerous problems arising in the physics of black holes.

 In this connection, the purpose of our paper is to derive, within
 the BV (section 2) and BLT (section~3) quantization methods for arbitrary
 gauge theories, the Ward identities for the generating functionals of
 Green's functions with external fields and to investigate the character
 of their dependence on the most general form of gauge fixing.

 We use the condensed notations introduced by De Witt \cite{11} and
 designations adopted in \cite{1,2}. The invariant tensor of the
 group $Sp(2)$ is denoted as $\varepsilon^{ab}$ ($a=1, 2$). It is a
 constant antisymmetric second rank tensor and subject to the
 normalization condition $\varepsilon^{12}=1$. The symmetrization of a
 certain quantity over the $Sp(2)$ indices is denoted as
 $A^{\{ab\}}=A^{ab}+A^{ba}$. Derivatives with respect to sources and
 antifields are understood as acting from the left, and those to fields,
 as acting from the
 right (unless otherwise specified); left derivatives
 with respect to the fields are labelled by the subscript ``${\it l}\,$"
 ($\delta_l/\delta\phi$ stands for the left derivative with
 respect to the field $\phi$).


\section{Quantum Gauge Theories with External Fields in the BV Formalism}
 Let us recall that the quantization of a gauge theory within the BV method
 \cite{1} (see also \cite{12}) implies introducing a complete set of
 fields $\phi^A$ and a set of the corresponding antifields $\phi^*_A$
 (which play the role of sources of the BRST transformations) with the
 following Grassmann parities:
\[
 \varepsilon(\phi^A)\equiv\varepsilon_A,\;\;
 \varepsilon(\phi^*_A)=\varepsilon_A+1.
\]
 The specific structure of the configuration space of the fields $\phi^A$
 (composed by the original classical fields, the (anti) ghost pyramids and
 the Lagrangian multipliers) is determined by the properties of the
 initial classical theory, i.e. by the linear dependence (for reducible
 theories) or independence (for irreducible theories) of the generators of
 gauge transformations.

 The extended generating functional $Z(J,\phi^*)$ of the Green functions
 for the fields of the complete configuration space is constructed within
 the BV quantization scheme by the rule
\begin{eqnarray}
 Z(J,\phi^*)&=&\int d\phi\;\exp\bigg\{\frac{i}{\hbar}\bigg(
 S_{\rm ext}(\phi,\phi^{*})+
 J_A\phi^A\bigg)\bigg\},
\end{eqnarray}
 where $\hbar$ is the Planck constant, $J_A$ are the usual sources
 to the fields $\phi^A$, $\varepsilon(J_A)=\varepsilon_A$,
 and $S_{\rm ext}=S_{\rm ext}(\phi,\phi^{*})$ is the gauge fixed quantum
 action defined by the relation
\begin{equation}
 \exp\bigg\{\frac{i}{\hbar}S_{\rm ext}\bigg\}=\exp\bigg(\hat{T}(\Psi)\bigg)
 \exp\bigg\{\frac{i}{\hbar}S\bigg\}.
\end{equation}
 Here, $S=S(\phi,\phi^*)$ is a bosonic functional satisfying the equation
\begin{equation}
 \frac{1}{2}(S,S)=i\hbar\Delta S,
\end{equation}
 or equivalently
\begin{equation}
 \Delta\exp\bigg\{\frac{i}{\hbar}S\bigg\}=0,
\end{equation}
 with the boundary condition ($\cal S$ is the initial gauge invariant
 classical action)
\begin{eqnarray*}
 S|_{\phi^{*}=\hbar=0}={\cal S}.
\end{eqnarray*}
 The operator $\hat{T}(\Psi)$ has the form
\begin{equation}
 \hat{T}(\Psi)=[\Delta,\Psi]_{+}\;,
\end{equation}
 where $\Psi$ is a fermionic functional fixing a specific choice of
 admissible gauge. In eqs.~(3)--(5) we use the standard definition of the
 antibracket $(\cdot,\cdot)$, given for two arbitrary functionals
 $F=F(\phi,\phi^*)$, $G=G(\phi,\phi^*)$ by the rule
\begin{eqnarray*}
 (F,\;G)=\frac{\delta F}{\delta\phi^A}\frac{\delta G}{\delta
 \phi^*_A}-(-1)^{(\varepsilon(F)+1)(\varepsilon(G)+1)}
 \frac{\delta G}{\delta\phi^A}\frac{\delta F}
 {\delta\phi^*_A}\,,
\end{eqnarray*}
 and the operator $\Delta$ is given by
\begin{eqnarray*}
 \Delta=(-1)^{\varepsilon_A}\frac{\delta_l}{\delta\phi^A}\frac
 {\delta}{\delta\phi^{*}_A}\,.
\end{eqnarray*}
 As a consequence of the nilpotency of ${\Delta}$, the relation
 $[\hat{T}(\Psi),\,\Delta]=0$ implies that the functional $S_{\rm ext}$ in
 (2) satisfies an equation of the same form as~(4)
\begin{eqnarray}
 {\Delta}\exp\bigg\{\frac{i}{\hbar}S_{\rm ext}\bigg\}=0.
\end{eqnarray}
 It is well-known that the gauge fixing (2) and (5) is in fact a particular
 case of the transformation generated by $\hat{T}(\Psi)$ with any fermionic
 operator chosen for $\Psi$ and describing the arbitrariness of solutions
 of eq.~(3). In this connection, we shall further deal with the most
 general case of gauge fixing, $\Psi$ being an arbitrary operator-valued
 functional.

 Let us consider the following representation of the generating
 functional $Z(J,\phi^*)$ in eq.~(1):
\begin{eqnarray}
 Z(J,\phi^*)&=&\int d\psi\;{\cal Z}({\cal J},\psi,\phi^*)\exp\bigg(\frac{i}
 {\hbar}{\cal Y}\psi\bigg),\nonumber\\
 \\
 {\cal Z}({\cal J},\psi,\phi^*)&=&\int d\varphi\;\exp\bigg\{\frac{i}
 {\hbar}\bigg(S_{\rm ext}(\varphi,\psi,\phi^{*})+{\cal J}\varphi\bigg)
 \bigg\}\nonumber
\end{eqnarray}
 with the decomposition
\[
 \phi^A=(\varphi^i,\;\psi^\alpha),\;\;\;
 J_A=({\cal J}_i,\;{\cal Y}_\alpha),\\
\]
\[
 \varepsilon(\varphi^i)\equiv\varepsilon_i,\;\;\;
 \varepsilon(\psi^\alpha)\equiv\varepsilon_\alpha.
\]
 In what follows we shall refer to ${\cal Z}({\cal J},\psi,\phi^*)$ in
 eq.~(7) as the extended generating functional of Green's functions with
 external fields $\psi^\alpha$. At the same time, given the set $\phi^A$,
 we impose no restrictions on the structure of the subset $\psi^\alpha$.

 In order to derive the Ward identities for a quantum gauge theory of
 general kind with external fields in the BV quantization approach we shall
 take advantage of the relation (6) for $S_{\rm ext}$. Indeed, multiplying
 eq.~(6) from the left by $\exp\{i/\hbar{\cal J}\varphi\}$ and integrating
 over the fields $\varphi^i$, we have
\begin{eqnarray}
 \int d\varphi\;\exp\bigg(\frac{i}{\hbar}{\cal J}_i\varphi^i\bigg)
 \Delta\exp\bigg\{\frac{i}{\hbar}S_{\rm ext}(\varphi,\psi,\phi^{*})
 \bigg\}=0.
\end{eqnarray}
 Then, using the equality
\begin{eqnarray}
 \exp\bigg\{\frac{i}{\hbar}{\cal J}_i\varphi^i\bigg\}
 \Delta=\bigg(\Delta-\frac{i}{\hbar}{\cal J}_i\frac{\delta}
 {\delta\varphi^{*}_i}\bigg)\exp\bigg\{\frac{i}{\hbar}{\cal J}_i\varphi^i
 \bigg\}
\end{eqnarray}
 and integrating in eq.~(8) by parts, we obtain for
 ${\cal Z}={\cal Z}({\cal J},\psi,\phi^*)$ the following Ward identity:
\begin{eqnarray}
 \hat{\omega}{\cal Z}=0,
\end{eqnarray}
 where $\hat{\omega}$ is the operator
\begin{eqnarray}
 \hat{\omega}=i\hbar\Delta_\psi+{\cal J}_i\frac{\delta}
 {\delta\varphi^{*}_i}\,,\;\;\;
 \Delta_\psi\equiv(-1)^{\varepsilon_\alpha}\frac{\delta_l}{\delta\psi^
 \alpha}\frac{\delta}{\delta\psi^{*}_\alpha}\,,
\end{eqnarray}
 possessing nilpotency $\hat{\omega}^2=0$.

 One readily establishes the fact that in terms of the generating
 functional of the connected Green functions with external fields
 ${\cal W}={\cal W}({\cal J},\psi,\phi^*)$ defined by
\[
 {\cal Z}=\exp\bigg\{\frac{i}{\hbar}{\cal W}\bigg\}
\]
 the identity~(10) takes on the form
\begin{eqnarray}
 \hat{\omega}{\cal W}=\frac{\delta{\cal W}}{\delta\psi^\alpha}
 \frac{\delta{\cal W}}{\delta\psi^*_\alpha}\,.
\end{eqnarray}
 Similarly, introducing the generating functional
 $\Gamma=\Gamma(\varphi,\psi,\phi^*)$ of the vertex Green functions with
 external fields
\[
 \Gamma(\varphi,\psi,\phi^*)={\cal W}({\cal J},\psi,\phi^*)-
 {\cal J}_i\varphi^i,\;\;\;
 \varphi^i=\frac{\delta{\cal W}}{\delta{\cal J}_i},\;\;\;
 {\cal J}_i=-\frac{\delta\Gamma}{\delta\varphi^i}\,,
\]
 we have, by virtue of eqs.~(11), (12) and the relations
\[
 \frac{\delta{\cal W}}{\delta\psi^\alpha}=\frac{\delta\Gamma}
 {\delta\psi^\alpha}\,,\;\;\;
 \frac{\delta{\cal W}}{\delta\phi^*_A}=\frac{\delta\Gamma}
 {\delta\phi^*_A}\,,
\]
 the following Ward identities:
\begin{equation}
 \frac{1}{2}(\Gamma,\Gamma)=i\hbar\Delta_\psi\Gamma-
 i\hbar(\Gamma^{''-1})^{ij}
 \bigg(\frac{\delta_l}{\delta\varphi^j}
 \frac{\delta\Gamma}{\delta\psi^\alpha}\bigg)
 \bigg(\frac{\delta}{\delta\psi^*_\alpha}
 \frac{\delta\Gamma}{\delta\varphi^i}\bigg),
\end{equation}
 where
\begin{eqnarray*}
 (\Gamma^{''-1})^{ik}(\Gamma^{''})_{kj}=\delta^i_j\,,\;\;\;
 (\Gamma^{''})_{ij}\equiv\frac{\delta_l}{\delta\varphi^i}
 \frac{\delta\Gamma}{\delta\varphi^j}\,.
\end{eqnarray*}

 Let us now investigate the gauge dependence of the above-introduced
 generating functionals with external fields under the most general
 variation
\[
 \delta\Psi\bigg(\phi^A,\,\phi^*_A;\,
 \frac{\delta_l}{\delta\phi^A},\,\frac{\delta}{\delta\phi^*_A}\bigg)=
 \delta\Psi\bigg(\varphi^i,\,\psi^\alpha,\,\phi^*_A;\,
 \frac{\delta_l}{\delta\varphi^i},\,\frac{\delta_l}{\delta\psi^\alpha},\,
 \frac{\delta}{\delta\phi^*_A}\bigg)
\]
 of the gauge fermion.

 From eq.~(2) it follows that the variation of
 $\exp\{i/\hbar S_{\rm ext}\}$ reads
\[
 \delta\bigg(\!\exp\bigg\{\frac{i}{\hbar}S_{\rm ext}\bigg\}\bigg)
 =\hat{T}(\delta X)\exp\bigg\{\frac{i}{\hbar}S_{\rm ext}\bigg\},
\]
 where $\delta X$ is related to $\delta\Psi$ through an operator-valued
 transformation (linear in $\delta\Psi$) whose explicit form is not
 essential for the following treatment. Note, however, that we can always
 dispose of the operator ordering so that $\delta X$ is represented as
\begin{eqnarray}
 \delta X\bigg(\varphi^i,\psi^\alpha,\phi^*_A;\,
 \frac{\delta_l}{\delta\varphi^i},\frac{\delta_l}{\delta\psi^\alpha},
 \frac{\delta}{\delta\phi^*_A}\bigg)=
 \delta X^{(0)}\bigg(\varphi^i,\psi^\alpha,\phi^*_A;\,
 \frac{\delta_l}{\delta\psi^\alpha},\frac{\delta}{\delta\phi^*_A}\bigg)
 \nonumber\\
 +
 \sum_{N=1}\frac{\delta_l}{\delta\varphi^{i_1}}\ldots
 \frac{\delta_l}{\delta\varphi^{i_N}}
 \delta X^{(i_1\ldots i_N)}\bigg(\varphi^i,\psi^\alpha,\phi^*_A;\,
 \frac{\delta_l}{\delta\psi^\alpha},\frac{\delta}{\delta\phi^*_A}\bigg).
\end{eqnarray}

 By virtue of eqs.~(5)--(7), the variation of the functional
 ${\cal Z}({\cal J},\psi,\phi^*)$ has the form
\begin{eqnarray}
 \delta{\cal Z}({\cal J},\psi,\phi^*)=
 \int d\varphi\;\exp\bigg(\frac{i}{\hbar}{\cal J}_i\varphi^i\bigg)\,
 \Delta\!\bigg(\delta X\exp\bigg\{\frac{i}{\hbar}
 S_{\rm ext}(\varphi,\psi,\phi^{*})\bigg\}\bigg).
\end{eqnarray}
 Then, using eq.~(9), definition~(11) and the representation~(14), we have,
 after the integration by parts in eq.~(15) has been performed,
\begin{eqnarray}
 \lefteqn{\delta{\cal Z}({\cal J},\psi,\phi^*)=}\nonumber\\
 & &\frac{1}{i\hbar}\hat{\omega}
 \delta X\bigg(\frac{\hbar}{i}\frac{\delta}{\delta{\cal J}_i},\,
 \psi^\alpha,\,\phi^*_A\,;(-1)^{\varepsilon_i}\frac{1}{i\hbar}{\cal J}_i,\,
 \frac{\delta_l}{\delta\psi^\alpha},\,\frac{\delta}{\delta\phi^*_A}\bigg)
 {\cal Z}({\cal J},\psi,\phi^*).
\end{eqnarray}
 One readily observes that in terms of the  generating functional
 ${\cal W}({\cal J},\psi,\phi^*)$ the relation~(16) takes on the form
\begin{eqnarray}
 \delta{\cal W}=-\hat{\Omega}\langle\delta X\rangle,
\end{eqnarray}
 where $\langle\delta X\rangle$ is the vacuum expectation value of the
  functional $\delta X$
\[
 \langle\delta X\rangle=\delta X
 \bigg(\frac{\delta{\cal W}}{\delta{\cal J}_i}+\frac{\hbar}{i}
 \frac{\delta}{\delta{\cal J}_i},\,\psi^\alpha,\,\phi^*_A\,;
 (-1)^{\varepsilon_i}\frac{1}{i\hbar}{\cal J}_i,\,
 \frac{i}{\hbar}\frac{\delta_l{\cal W}}{\delta\psi^\alpha}+
 \frac{\delta_l}{\delta\psi^\alpha}\,,
 \frac{i}{\hbar}\frac{\delta{\cal W}}{\delta\phi^*_A}+
 \frac{\delta}{\delta\phi^*_A}\bigg),
\]
 and $\hat{\Omega}$ is an operator defined according to
\[
 \hat{\Omega}=\exp\bigg\{\!\!-\frac{i}{\hbar}{\cal W}\bigg\}\hat{\omega}
 \exp\bigg\{\frac{i}{\hbar}{\cal W}\bigg\}.
\]
 By virtue of the Ward identities (12) for the functional
 ${\cal W}({\cal J},\psi,\phi^*)$, the operator $\hat{\Omega}$ can be
 represented as
\begin{eqnarray}
 \hat{\Omega}=\hat{\omega}-\frac{\delta{\cal W}}{\delta\psi^\alpha}
 \frac{\delta}{\delta\psi^*_\alpha}-(-1)^{\varepsilon_\alpha}
 \frac{\delta{\cal W}}{\delta\psi^*_\alpha}\frac{\delta_l}
 {\delta\psi^\alpha}\,,
\end{eqnarray}
 while from the nilpotency of $\hat{\omega}$ it follows that
\begin{eqnarray}
 \hat{\Omega}^2=0.
\end{eqnarray}

 In order to determine the character of the gauge dependence of
 $\Gamma(\varphi,\psi,\phi^*)$ we first observe that
 $\delta\Gamma=\delta{\cal W}$ and, secondly,
\begin{eqnarray*}
 \left.\frac{\delta}{\delta{\cal J}_i}\right|_{\psi,\;\phi^*}&=&
 \left.\frac{\delta\varphi^j}{\delta{\cal J}_i}
 \frac{\delta_l}{\delta\varphi^j}\right|_{\psi,\;\phi^*},\\
 \left.\frac{\delta_l}{\delta\psi^\alpha}\right|_{{\cal J},\;\phi^*}&=&
 \left.\frac{\delta_l}{\delta\psi^\alpha}\right|_{\varphi,\;\phi^*}+
 \left.\frac{\delta_l\varphi^i}{\delta\psi^\alpha}
 \frac{\delta_l}{\delta\varphi^i}\right|_{\psi,\;\phi^*},\\
 \left.\frac{\delta}{\delta\phi^*_A}\right|_{{\cal J},\;\psi}&=&
 \left.\frac{\delta}{\delta\phi^*_A}\right|_{\varphi,\;\psi}+
 \left.\frac{\delta\varphi^i}{\delta\phi^*_A}
 \frac{\delta_l}{\delta\varphi^i}\right|_{\psi,\;\phi^*}.
\end{eqnarray*}
 At the same time, note that
\begin{eqnarray*}
 \frac{\delta\varphi^j}{\delta{\cal J}_i}&=&-(\Gamma^{''-1})^{ij},\\
 \frac{\delta_l\varphi^i}{\delta\psi^\alpha}&=&-(-1)^
 {\varepsilon_i\varepsilon_\alpha}(\Gamma^{''-1})^{ij}
 \bigg(\frac{\delta_l}{\delta\varphi^j}
 \frac{\delta_l\Gamma}{\delta\psi^\alpha}\bigg)\,,
 \\
 \frac{\delta\varphi^i}{\delta\phi^*_A}&=&-(-1)^
 {\varepsilon_i(\varepsilon_A+1)}(\Gamma^{''-1})^{ij}
 \bigg(\frac{\delta_l}{\delta\varphi^j}
 \frac{\delta\Gamma}{\delta\phi^*_A}\bigg)\,,
\end{eqnarray*}
 where
\begin{eqnarray*}
 \frac{\delta_l{\cal J}_j}{\delta\varphi^i}=-(\Gamma^{''})_{ij}.
\end{eqnarray*}

 By virtue of eq.~(17) and the above equations, the variation of
 $\Gamma(\varphi,\psi,\phi^*)$ reads
\begin{eqnarray}
 \delta\Gamma=-\hat{q}\langle\langle\delta X\rangle\rangle.
\end{eqnarray}
 Here,
\[
 \langle\langle\delta X\rangle\rangle
 =\delta X\bigg(\varphi^i+
 i\hbar(\Gamma^{''-1})^{ij}
 \frac{\delta_l}{\delta\varphi^j},\;\psi^\alpha,\;
 \phi^*_A;\;\frac{i}{\hbar}\frac{\delta_l\Gamma}{\delta\varphi_i},
 \rule{4cm}{0cm}
\]
\[
 \rule{.6cm}{0cm} \frac{\delta_l}{\delta\psi^\alpha}
 +\frac{i}{\hbar}\frac{\delta_l\Gamma}
 {\delta\psi^\alpha}-(-1)^{\varepsilon_i\;\varepsilon_\alpha}\;\cdot\;
 (\Gamma^{''-1})^{ij}
 \bigg(\frac{\delta_l}{\delta\varphi^j}
 \frac{\delta_l\Gamma}{\delta\psi^\alpha}\bigg)
 \frac{\delta_l}{\delta\varphi^i},
\]
\[
 \rule{.9cm}{0cm} \frac{\delta}{\delta\phi^*_A}+
 \frac{i}{\hbar}\frac{\delta\Gamma}{\delta\phi^*_A}
 -(-1)^{\varepsilon_i(\varepsilon_A+1)}(\Gamma^{''-1})^{ij}
 \bigg(\frac{\delta_l}{\delta\varphi^j}\frac{\delta\Gamma}{\delta\phi^*_A}
 \bigg)\frac{\delta_l}{\delta\varphi^i}\bigg),
\]
 and $\hat{q}$ is an operator of the form
\[
 \hat{q}=i\hbar(-1)^{\varepsilon_\alpha}\bigg(
 \frac{\delta_l}{\delta\psi^\alpha}-
 (-1)^{\varepsilon_i\;\varepsilon_\alpha}\;\cdot\;\;
 (\Gamma^{''-1})^{ij}\;\,
 \bigg(\frac{\delta_l}{\delta\varphi^j}
 \frac{\delta_l\Gamma}{\delta\psi^\alpha}\bigg)
 \frac{\delta_l}{\delta\varphi^i}\bigg)\cdot \rule{2.1cm}{0cm}
\]
\[
 \cdot\bigg(\frac{\delta}{\delta\psi^*_\alpha}
 -(-1)^{\varepsilon_m(\varepsilon_\alpha+1)}(\Gamma^{''-1})^{mn}
 \bigg(\frac{\delta_l}{\delta\varphi^n}
 \frac{\delta\Gamma}{\delta\psi^*_\alpha}\bigg)
 \frac{\delta_l}{\delta\varphi^m}\bigg)
\]
\[
 -\frac{\delta\Gamma}{\delta\phi^A}\bigg(
 \frac{\delta}{\delta\phi^*_A}
 -(-1)^{\varepsilon_i\,(\varepsilon_A+1)}\,(\Gamma^{''-1})^{ij}\;\;
 \bigg(\frac{\delta_l}{\delta\varphi^j}\frac{\delta\Gamma}{\delta\phi^*_A}
 \bigg)\frac{\delta_l}{\delta\varphi^i}\bigg) \rule{1.2cm}{0cm}
\]
\begin{equation}
 -(-1)^{\varepsilon_\alpha}
 \frac{\delta\Gamma}{\delta\psi^*_\alpha}\bigg(
 \frac{\delta_l}{\delta\psi^\alpha}
 -(-1)^{\varepsilon_i\;\varepsilon_\alpha}\;\cdot\;\;(\Gamma^{''-1})^{ij}
\;\;
 \bigg(\frac{\delta_l}{\delta\varphi^j}
 \frac{\delta_l\Gamma}{\delta\psi^\alpha}\bigg)
 \frac{\delta_l}{\delta\varphi^i}\bigg). \rule{1.5cm}{0cm}
\end{equation}
 Since $\hat{q}$ is related to $\hat{\Omega}$ in (18) through the Legendre
 transformation, it also possesses the property of nilpotency (cf.
eq.~(19))
\[
 \hat{q}^2=0.
\]


\section{Theories with External Fields in the BLT Scheme}
 We now consider the structure of quantum gauge theories with external
 fields within the BLT quantization method \cite{2}. To this end, we
 remind that the variables of the formalism in \cite{2} are composed by
 fields $\phi^A$ and a set of the corresponding antifields
 $\phi^*_{Aa}$, $\bar{\phi}_A$ (the doublets of antifields $\phi^*_{Aa}$
 play the role of sources of the BRST and antiBRST transformations, while
 the antifields $\bar{\phi}_A$ are the sources of the mixed BRST and
 antiBRST transformations) with
\[
 \varepsilon(\phi^A)=\varepsilon_A,\;\;
 \varepsilon(\phi^*_{Aa})=\varepsilon_A+1,\;\;
 \varepsilon(\bar{\phi}_A)=\varepsilon_A.
\]
 Note that for any given gauge theory the structure of the configuration
 space of the fields $\phi^A$ in the BLT formalism coincides with that
 of the BV method. At the same time, the fields in the BLT scheme form
 components of irreducible completely symmetric $Sp(2)$-tensors.

 The extended generating functional $Z(J,\phi^*,\bar{\phi})$ of Green's
 functions for the fields of the complete configuration space is defined
 in the quantization scheme \cite{2} by the rule
\begin{eqnarray}
 Z(J,\phi^*,\bar{\phi})&=&\int d\phi\;\exp\bigg\{\frac{i}{\hbar}\bigg(
 S_{\rm ext}(\phi,\phi^{*},\bar{\phi})+J_A\phi^A\bigg)\bigg\},
\end{eqnarray}
 where $S_{\rm ext}=S_{\rm ext}(\phi,\phi^{*},\bar{\phi})$ is the gauge
 fixed quantum action
\begin{equation}
 \exp\bigg\{\frac{i}{\hbar}S_{\rm ext}\bigg\}=\exp\bigg(\!\!-i\hbar
 \hat{T}(F)\bigg)\exp\bigg\{\frac{i}{\hbar}S\bigg\}.
\end{equation}
 Here, $S=S(\phi,\phi^*,\bar{\phi})$ is a bosonic functional satisfying
 the equations
\begin{eqnarray}
 \frac{1}{2}(S,S)^a+V^aS=i\hbar\Delta^aS,
\end{eqnarray}
 or equivalently
\begin{eqnarray}
 \bar{\Delta}^a\exp\bigg\{\frac{i}{\hbar}S\bigg\}=0,\;\;\;
 \bar{\Delta}^a\equiv\Delta^a+\frac{i}{\hbar}V^a,
\end{eqnarray}
 with the boundary condition (${\cal S}$ is the classical action)
\begin{eqnarray*}
 S|_{\phi^{*}=\bar{\phi}=\hbar=0}={\cal S}.
\end{eqnarray*}
 $\hat{T}(F)$ is an operator defined by
\begin{equation}
 \hat{T}(F)=\frac{1}{2}\varepsilon_{ab}[\bar{\Delta}^b,[\bar{\Delta}^a,
 F]_{-}]_{+}\,,
\end{equation}
 where $F$ is a bosonic (generally, operator-valued) functional fixing a
 specific choice of gauge. In eqs.~(24)--(26) we use the definition
 \cite{2} of the extended antibrackets $(\cdot,\cdot)^a$, introduced for
 two arbitrary functionals $F=F(\phi,\phi^*,\bar{\phi})$ and
 $G=G(\phi,\phi^*,\bar{\phi})$ by the rule
\begin{eqnarray*}
 (F,G)^a=\frac{\delta F}{\delta\phi^A}\frac{\delta G}{\delta
 \phi^{*}_{Aa}}-
 (-1)^{(\varepsilon(F)+1)(\varepsilon(G)+1)}
 \frac{\delta G}{\delta\phi^A}\frac{\delta F} {\delta\phi^{*}_{Aa}},
\end{eqnarray*}
 and the operators $\Delta^a$ and $V^a$
\begin{eqnarray*}
 \Delta^a=(-1)^{\varepsilon_A}\frac{\delta_l}{\delta\phi^A}\frac
 {\delta}{\delta\phi^{*}_{Aa}}\;,\;\;
 V^a=\varepsilon^{ab}\phi^{*}_{Ab}\frac{\delta}{\delta\bar\phi_A}\;.
\end{eqnarray*}
 As a consequence of the algebraic properties of $\bar{\Delta}^a$ in (25),
 i.e. $\bar{\Delta}^{\{a}\bar{\Delta}^{b\}}=0$, the relations
 $[\hat{T}(F),\,\bar{\Delta}^a]=0$ imply that the functional
 $S_{\rm ext}$ in (23) satisfies the equations
\begin{eqnarray}
 \bar{\Delta}^a\exp\bigg\{\frac{i}{\hbar}S_{\rm ext}\bigg\}=0.
\end{eqnarray}

 Let us now introduce the following representation of the generating
 functional $Z(\phi,\phi^*,\bar{\phi})$ in eq.~(22):
\begin{eqnarray*}
 Z(J,\phi^*,\bar{\phi})=\int d\psi\;{\cal Z}
 ({\cal J},\psi,\phi^*,\bar{\phi})
 \exp\bigg(\frac{i}{\hbar}{\cal Y}\psi\bigg),
\end{eqnarray*}
 where ${\cal Z}({\cal J},\psi,\phi^*,\bar{\phi})$ denotes the extended
 generating functional of Green's functions with the external fields
 $\psi^\alpha$
\begin{eqnarray}
 {\cal Z}({\cal J},\psi,\phi^*,\bar{\phi})=\int d\varphi\;
 \exp\bigg\{\frac{i}
 {\hbar}\bigg(S_{\rm ext}(\varphi,\psi,\phi^{*},\bar{\phi})+
 {\cal J}\varphi\bigg)\bigg\}.
\end{eqnarray}
 Here, as before, we abbreviate $\phi^A=(\varphi^i,\;\psi^\alpha)$ and
 $J_A=({\cal J}_i,\;{\cal Y}_\alpha)$.

 As in the case of the BV quantization scheme, the Ward identities
 for a quantum gauge theory of general kind with external fields considered
 in the BLT method follow immediately from the equation for the gauge
 fixed quantum action $S_{\rm ext}$. Eq.~(27) implies
\begin{eqnarray}
 \int d\varphi\;\exp\bigg(\frac{i}{\hbar}{\cal J}_i\varphi^i\bigg)
 \bar{\Delta}^a\exp\bigg\{\frac{i}{\hbar}S_{\rm ext}(\varphi,\psi,\phi^{*},
 \bar{\phi})\bigg\}=0.
\end{eqnarray}
 Then, taking into account the equalities
\begin{eqnarray}
 \exp\bigg\{\frac{i}{\hbar}{\cal J}_i\varphi^i\bigg\}
 \bar{\Delta}^a=\bigg(\bar{\Delta}^a-\frac{i}{\hbar}{\cal J}_i\frac{\delta}
 {\delta\varphi^{*}_{ia}}\bigg)\exp\bigg
 \{\frac{i}{\hbar}{\cal J}_i\varphi^i \bigg\}
\end{eqnarray}
 and performing the integration by parts in eq.~(29), we have
\begin{eqnarray}
 \hat{\omega}^a{\cal Z}=0,
\end{eqnarray}
 where $\hat{\omega}^a$ stand for the operators
\begin{eqnarray}
 \hat{\omega}^a=i\hbar{\Delta}^a_\psi+{\cal J}_i\frac{\delta}
 {\delta\varphi^{*}_{ia}}-V^a\,,\;\;\;
 {\Delta}^a_\psi\equiv(-1)^{\varepsilon_\alpha}\frac{\delta_l}
 {\delta\psi^\alpha}\frac{\delta}{\delta\psi^{*}_{\alpha a}}
\end{eqnarray}
 with the properties
\begin{eqnarray}
 \hat{\omega}^{\{a}\hat{\omega}^{b\}}=0.
\end{eqnarray}
 The relations (31) together with (32) determine the Ward identities
 for the generating functional ${\cal Z} ({\cal J},\psi,\phi^*,\bar{\phi})$
 in (28).

 Rewriting eqs.~(31) in terms of the generating functionals ${\cal W}$,
 with ${\cal Z}=\exp\{i/\hbar{\cal W}\}$, and $\Gamma$, where
\[
 \Gamma(\varphi,\psi,\phi^*,\bar{\phi})=
 {\cal W}({\cal J},\psi,\phi^*,\bar{\phi})-{\cal J}_i\varphi^i,\;\;\;
 \varphi^i=\frac{\delta{\cal W}}{\delta{\cal J}_i}\,,\;\;\;
 {\cal J}_i=-\frac{\delta\Gamma}{\delta\varphi^i}\,,
\]
\[
 \frac{\delta{\cal W}}{\delta\psi^\alpha}=\frac{\delta\Gamma}
 {\delta\psi^\alpha}\,,\;\;\;
 \frac{\delta{\cal W}}{\delta\phi^*_{Aa}}=\frac{\delta\Gamma}
 {\delta\phi^*_{Aa}}\,,\;\;\;
 \frac{\delta{\cal W}}{\delta\bar{\phi}_A}=\frac{\delta\Gamma}
 {\delta\bar{\phi}_A}\,,
\]
 we recast the Ward identities into the form
\begin{eqnarray}
 \hat{\omega}^a{\cal W}=\frac{\delta{\cal W}}{\delta\psi^\alpha}
 \frac{\delta{\cal W}}{\delta\psi^*_{\alpha a}}\,,
\end{eqnarray}
\begin{equation}
 \frac{1}{2}(\Gamma,\Gamma)^a+V^a\Gamma=i\hbar\Delta_\psi^a\Gamma
 -i\hbar(\Gamma^{''-1})^{ij}
 \bigg(\frac{\delta_l}{\delta\varphi^j}
 \frac{\delta\Gamma}{\delta\psi^\alpha}\bigg)
 \bigg(\frac{\delta}{\delta\psi^*_{\alpha a}}
 \frac{\delta\Gamma}{\delta\varphi^i}\bigg).
\end{equation}

 We shall now study the change of the above-introduced generating
 functionals ${\cal Z}$, ${\cal W}$, $\Gamma$ in the BLT quantization
 scheme under the variation of the gauge boson $F$, chosen in the
 most general form of an operator-valued functional, i.e.
\[
 \delta\!  F\!\bigg(\!\!\phi^A,\phi^*_{Aa},\bar{\phi}_A;\,
 \frac{\delta_l}{\delta\phi^A},\frac{\delta}{\delta\phi^*_{Aa}},
 \frac{\delta}{\delta\bar{\phi}_A}\!\bigg) \!\!\!\;=\!
 \delta\!  F\!\bigg(\!\!\varphi^i,\psi^\alpha,\phi^*_{Aa},\bar{\phi}_A;\,
 \frac{\delta_l}{\delta\varphi^i},\frac{\delta_l}{\delta\psi^\alpha},
 \frac{\delta}{\delta\phi^*_{Aa}},\frac{\delta}{\delta\bar{\phi}_A}\!\bigg).
\]
 Taking eq.~(23) into account, we have
\[
 \delta\bigg(\!\exp\bigg\{\frac{i}{\hbar}S_{\rm ext}\bigg\}\bigg)
 =-i\hbar\hat{T}(\delta Y)\exp\bigg\{\frac{i}{\hbar}S_{\rm ext}\bigg\}
\]
 and consequently, with eqs.~(25) and (26),
\begin{eqnarray}
 \lefteqn{\delta{\cal Z}({\cal J},\psi,\phi^*,\bar{\phi})=} \nonumber\\
 & &\frac{i\hbar}{2}
 \varepsilon_{ab}\int d\varphi\;\exp\bigg(\frac{i}{\hbar}
 {\cal J}_i\varphi^i\bigg)\bar{\Delta}^a\bar{\Delta}^b\bigg(\delta Y
 \exp\bigg\{\frac{i}{\hbar}S_{\rm ext}(\varphi,\psi,\phi^{*},\bar{\phi})
 \bigg\}\bigg),
\end{eqnarray}
 where $\delta Y$ (related to $\delta F$ through a linear
 transformation) can always be represented as
\begin{eqnarray}
 \delta  Y\bigg(\varphi^i,\psi^\alpha,\phi^*_{Aa},\bar{\phi}_A;\,
 \frac{\delta_l}{\delta\varphi^i},\frac{\delta_l}{\delta\psi^\alpha},
 \frac{\delta}{\delta\phi^*_{Aa}},\frac{\delta}{\delta\bar{\phi}_A}\bigg)&=&
\nonumber\\
 \delta Y^{(0)}\bigg(\varphi^i,\psi^\alpha,\phi^*_{Aa}\!\!\!\!\!\!&,&
 \!\!\!\! \bar{\phi}_A;\,
 \frac{\delta_l}{\delta\psi^\alpha},\frac{\delta}{\delta\phi^*_{Aa}},
 \frac{\delta}{\delta\bar{\phi}_A}\bigg)                              \\
 +
 \sum_{N=1}\frac{\delta_l}{\delta\varphi^{i_1}}\ldots
 \frac{\delta_l}{\delta\varphi^{i_N}}
 \delta Y^{(i_1\ldots
i_N)}\bigg(\varphi^i,\psi^\alpha,\,\phi^*_{Aa}\!\!\!\!\!\!&,&\!\!\!\!
 \bar{\phi}_A;\,\frac{\delta_l}{\delta\psi^\alpha},
 \frac{\delta}{\delta\phi^*_{Aa}},\frac{\delta}{\delta\bar{\phi}_A}\bigg)
\nonumber.
\end{eqnarray}

 By virtue of eqs.~(30), (32), (37) and integrating by parts in eq.~(36)
 one finds
\[
 \delta{\cal Z}({\cal J},\psi,\phi^*,\bar{\phi})=\frac{i}{2\hbar}
 \varepsilon_{ab}\hat{\omega}^b\hat{\omega}^a\cdot
\]
\begin{equation}
 \cdot\delta Y\!\bigg(\frac{\hbar}{i}\frac{\delta}{\delta{\cal J}_i},\,
 \psi^\alpha,\phi^*_{Aa},\bar{\phi}_A;\,(-1)^{\varepsilon_i}
 \frac{1}{i\hbar}{\cal J}_i,\,
 \frac{\delta_l}{\delta\psi^\alpha},\,\frac{\delta_l}{\delta\phi^*_{Aa}},\,
 \frac{\delta_l}{\delta\bar{\phi}_A}\bigg)
 {\cal Z}\!({\cal J},\psi,\phi^*,\bar{\phi}).
\end{equation}
 Given this and using arguments quite similar to those presented in the
 case of the BV scheme, one readily establishes the fact that in terms of
 the generating functionals ${\cal W} ({\cal J},\psi,\phi^*,\bar{\phi})$
 and $\Gamma(\varphi,\psi,\phi^*,\bar{\phi})$ the corresponding variations
 have the form
\begin{eqnarray}
 \delta{\cal W}=\frac{1}{2}\varepsilon_{ab}\hat{\Omega}^b\hat{\Omega}^a
 \langle\delta Y\rangle,
\end{eqnarray}
\begin{eqnarray}
 \delta\Gamma=\frac{1}{2}\varepsilon_{ab}\hat{q}^b\hat{q}^a
 \langle\langle\delta Y\rangle\rangle,
\end{eqnarray}
 where
\[
 \langle\delta Y\rangle= \delta Y
 \bigg(\frac{\delta{\cal W}}{\delta{\cal J}_i}+\frac{\hbar}{i}
 \frac{\delta}{\delta{\cal J}_i},\,\psi^\alpha,\,\phi^*_{Aa},\,
 \bar{\phi}_A\,;
 (-1)^{\varepsilon_i}\frac{1}{i\hbar}{\cal J}_i,\,
\]
\[
 \rule{3.5cm}{0cm}
 \frac{i}{\hbar}\frac{\delta_l{\cal W}}{\delta\psi^\alpha}+
 \frac{\delta_l}{\delta\psi^\alpha}\,,
 \frac{i}{\hbar}\frac{\delta{\cal W}}{\delta\phi^*_{Aa}}+
 \frac{\delta}{\delta\phi^*_{Aa}}\,,
 \frac{i}{\hbar}\frac{\delta{\cal W}}{\delta\bar{\phi}_A}+
 \frac{\delta}{\delta\bar{\phi}_A}\bigg)
\]
 and
\[
 \langle\langle\delta Y\rangle\rangle=
 \delta Y\bigg(\varphi^i+i\hbar(\Gamma^{''-1})^{ij}
 \frac{\delta_l}{\delta\varphi^j},\psi^\alpha,
 \phi^*_{Aa},\bar{\phi}_A;\;
 \frac{i}{\hbar}\frac{\delta_l\Gamma}{\delta\varphi_i},\rule{4cm}{0cm}
\]
\[
\begin{array}{r@{}l@{}l@{}l}
 \displaystyle \rule{2cm}{0cm}
 \frac{\delta_l}{\delta\psi^\alpha}
 &\displaystyle
 +\frac{i}{\hbar}\frac{\delta_l\Gamma}
 {\delta\psi^\alpha}
 &\displaystyle
  -(-1)^{\varepsilon_i\;\varepsilon_\alpha}\;\cdot\;\;
 &\displaystyle
 (\Gamma^{''-1})^{ij}
 \bigg(\frac{\delta_l}{\delta\varphi^j}
 \frac{\delta_l\Gamma}{\delta\psi^\alpha}\bigg)
 \frac{\delta_l}{\delta\varphi^i}\,,                         \\[3ex]
 \displaystyle
 \frac{\delta}{\delta\phi^*_{Aa}}
 &\displaystyle
 +\frac{i}{\hbar}\frac{\delta\Gamma}{\delta\phi^*_{Aa}}
 &\displaystyle
-(-1)^{\varepsilon_i(\varepsilon_A+1)}
 &\displaystyle
 (\Gamma^{''-1})^{ij}
 \bigg(\frac{\delta_l}{\delta\varphi^j}
 \frac{\delta\Gamma}{\delta\phi^*_{Aa}}\bigg)
 \frac{\delta_l}{\delta\varphi^i}\,,                         \\[3ex]
 \displaystyle
 \frac{\delta}{\delta\bar{\phi}_A}
 &\displaystyle
 +\frac{i}{\hbar}\frac{\delta\Gamma}{\delta\bar{\phi}_A}
 &\displaystyle
 -(-1)^{\varepsilon_i\;\varepsilon_A}\;\cdot\;
 &\displaystyle
 (\Gamma^{''-1})^{ij}
 \bigg(\frac{\delta_l}{\delta\varphi^j}
 \frac{\delta\Gamma}{\delta\bar{\phi}_A}\bigg)
 \frac{\delta_l}{\delta\varphi^i}\;\bigg).
\end{array}
\]
 In eq.~(39), $\hat{\Omega}^a$ stand for the operators
\[
 \hat{\Omega}^a=\exp\bigg\{\!\!-\frac{i}{\hbar}{\cal W}\bigg\}
 \hat{\omega}^a\exp\bigg\{\frac{i}{\hbar}{\cal W}\bigg\},
\]
 whose explicit form is
\begin{eqnarray}
 \hat{\Omega}^a=\hat{\omega}^a-\frac{\delta{\cal W}}{\delta\psi^\alpha}
 \frac{\delta}{\delta\psi^*_{\alpha a}}-(-1)^{\varepsilon_\alpha}
 \frac{\delta{\cal W}}{\delta\psi^*_{\alpha a}}\frac{\delta_l}
 {\delta\psi^\alpha}\,.
\end{eqnarray}
 Taking into account eq.~(33) and the Ward identities (34) for
 ${\cal W}({\cal J},\psi,\phi^*,\bar{\phi})$, we have
\begin{eqnarray}
 \hat{\Omega}^{\{a}\hat{\Omega}^{b\}}=0.
\end{eqnarray}
 At the same time, $\hat{q}^a$ in eq.~(40) are the operators
\begin{eqnarray}
 \hat{q}^a=i\hbar(-1)^{\varepsilon_\alpha}\bigg(
 \frac{\delta_l}{\delta\psi^\alpha}
 \!\!&-&\!\!(-1)^{\varepsilon_i\;\varepsilon_\alpha}\;\cdot\;\;
 (\Gamma^{''-1})^{ij}\;\:
 \bigg(\frac{\delta_l}{\delta\varphi^j}
 \frac{\delta_l\Gamma}{\delta\psi^\alpha}\bigg)
 \frac{\delta_l}
 {\delta\varphi^i}\bigg)\cdot                               \nonumber\\
 \cdot\bigg(\frac{\delta}{\delta\psi^*_{\alpha a}}
 \!\!&-&\!\!(-1)^{\varepsilon_m(\varepsilon_\alpha+1)}(\Gamma^{''-1})^{mn}
\bigg(
 \frac{\delta_l}{\delta\varphi^n}\frac{\delta\Gamma}
 {\delta\psi^*_{\alpha a}}\bigg)
 \frac{\delta_l}{\delta\varphi^m}\bigg)                     \nonumber\\
 -\frac{\delta\Gamma}{\delta\phi^A}\bigg(
 \frac{\delta}{\delta\phi^*_{Aa}}
 \!\!&-&\!\!(-1)^{\varepsilon_i(\varepsilon_A+1)}\;(\Gamma^{''-1})^{ij}\;\:
\bigg(
 \frac{\delta_l}{\delta\varphi^j}\frac{\delta\Gamma}{\delta\phi^*_{Aa}}
 \bigg)\frac{\delta_l}{\delta\varphi^i}\bigg)                       \\
 -(-1)^{\varepsilon_\alpha}
 \frac{\delta\Gamma}{\delta\psi^*_{\alpha a}}\bigg(
 \frac{\delta_l}{\delta\psi^\alpha}
 \!\!&-&\!\!(-1)^{\varepsilon_i\;\varepsilon_\alpha}\;\cdot\;\;
 (\Gamma^{''-1})^{ij}\;\:
 \bigg(\frac{\delta_l}{\delta\varphi^j}
 \frac{\delta_l\Gamma}{\delta\psi^\alpha}\bigg)
 \frac{\delta_l}{\delta\varphi^i}\bigg)                     \nonumber\\
 -\varepsilon^{ab}\phi^*_{Ab}\bigg(
 \frac{\delta}{\delta\bar{\phi}_A}
 \!\!&-&\!\!(-1)^{\varepsilon_i\;\varepsilon_A}\;\cdot\;\;
 (\Gamma^{''-1})^{ij}\;\:\bigg(
 \frac{\delta_l}{\delta\varphi^j}\frac{\delta\Gamma}{\delta\bar{\phi}_A}
 \bigg)\frac{\delta_l}{\delta\varphi^i}\bigg)              \nonumber
\end{eqnarray}
 related to $\hat{\Omega}^a$ through the Legendre transformation and,
 consequently, also possessing the properties (cf. eq.~(42))
\begin{eqnarray*}
 \hat{q}^{\{a}\hat{q}^{b\}}=0.
\end{eqnarray*}


\section{Conclusion}
 In this paper we have considered generating functionals of Green's
 functions with external fields in the framework of BV \cite{1}
 and BLT \cite{2} quantization schemes for general gauge theories.
 Note that, as compared to the study of \cite{13}, our approach
 closely follows the standard prescriptions of \cite{1,2} without
 having recourse to introducing an additional composite operator to the
 path integral.

 The Ward identities for the generating functionals with external fields
 are obtained: (10), (12), (13), (31), (34) and (35). The explicit gauge
 dependence on the most general form of gauge fixing has been derived: (16),
 (17), (20) and (38)--(40). In this connection, it is a quite
 remarkable fact that the gauge dependence concerned is described
 with the help of
 nilpotent fermionic operators (11), (18) and (21) in the BV formalism and
 doublets (32), (41) and (43) of nilpotent anticommuting operators in the
 BLT method, as in the case of quantum gauge theories of general
 kind with composite fields (see \cite{7}).

 The use of external or background fields plays an important role if
 a (background field) gauge invariant effective action is to be considered.
 A recently discussed approach
 is the method of the effective average action \cite{14} which allows to
 derive renormalization group equations for the coupling constants of a
 theory in a way which goes behind the standard one-loop calculation.
 However, by
 the present work, the way for using such methods in theories requiring BV
 or BLT quantization, e.~g. string theories, is not yet prepared.
 Namely, if a general gauge theory has to be considered where the
 fields are to be decomposed into a quantized part $\varphi^A$ and a
 classical background configuration $\phi^A_{\rm cl}$,
 i.e. $\Phi^A = \varphi^A + \phi^A_{\rm cl}$, the present method has to
 be generalized. Of course, such an approach will be of interest if
 vacuum condensates or topological nontrivial configurations like
 instantons --- as in the case of QCD --- play an essential role. This
 question has been considered independently in a recent paper \cite{15}.
 The extension of that method to the BV and BLT approach will be
 considered later on.


\section*{Acknowledgments}
 The authors would like to thank S.D. Odintsov, D.~M\"ulsch and
 A.A.~Reshetnyak for useful discussions. One of the authors (PL) is
 grateful for kind hospitality extended to him by Institut f\"{u}r
 Theoretische Physik of Hannover University and by Graduate College
 "Quantum Field Theory" at Center of Theoretical Sciences
  (NTZ) of Leipzig University during the summer half, 1996 -- 1997.

 The work has been partially supported by the Russian Foundation for
 Basic Research (RFBR), project 96--02--16017, as well as by the joint
 project of Deutsche Forschungsgemeinschaft and Russian Foundation for
 Basic Research (DFG--RFBR), 96--02--00180G.


\end{document}